\documentclass[useAMS]{mn2e}

\newif\ifAMStwofonts

\usepackage{graphicx}

\title[Reflection, absorption and X-ray polarimetry -- Predictions for MCG-6-30-15]
      {X-ray polarimetry as a new tool to discriminate reflection from absorption scenarios -- Predictions for MCG-6-30-15}
\author[F. Marin et al.]
      {F.~Marin$^1$\thanks{frederic.marin@astro.unistra.fr},
       R. W.~Goosmann$^1$, M.~Dov{\v c}iak$^2$, F. Muleri$^3$, D. Porquet$^1$,
       N. Grosso$^1$,
       \newauthor 
       V. Karas$^2$, and G. Matt$^4$ \\
       $^1$Observatoire Astronomique de Strasbourg, Universit\'e de Strasbourg, 
       CNRS, UMR 7550, 11 rue de l'Universit\'e, 67000 Strasbourg, France \\
       $^2$Astronomical Institute of the Academy of Sciences, Bo{\v c}ni
       II 1401, 14131 Prague, Czech Republic \\
       $^3$INAF/IAPS, Via del Fosso del Cavaliere 100, I-00133 Roma, Italy \\
       $^4$Dipartimento di Fisica, Universit\'a degli Studi Roma Tre, 
       Via della Vasca Navale 84, I-00146 Roma, Italy}
\date{Accepted 2012 August 13.
      Received 2012 August 13;
      in original form 2012 July 14}

\pagerange{\pageref{firstpage}--\pageref{lastpage}}
\pubyear{2012}

\begin{document}

\maketitle

\label{firstpage}

\begin{abstract}
We present modelling of X-ray polarisation spectra emerging from the
two competing scenarios that are proposed to explain the broad Fe
K$\alpha$ line in the Seyfert 1 galaxy MCG-6-30-15. The polarisation 
signature of complex absorption is studied for a partial covering 
scenario using a clumpy wind and compared to a reflection model based 
on the lamp-post geometry. The shape of the polarisation percentage 
and angle as a function of photon energy are found to be distinctly 
different between the reflection and the absorption case. Relativistic 
reflection produces significantly stronger polarisation in the 1--10~keV 
energy band than absorption. The spectrum of the polarisation angle adds 
additional constraints: in the absorption case it shows a constant shape, 
whereas the relativistic reflection scenario typically leads to a smooth
rotation of the polarisation angle with photon energy. Based on this
work, we conclude that a soft X-ray polarimeter on-board a small X-ray
satellite may already discriminate between the absorption and the
reflection scenario. A promising opportunity may arise 
with the {\it X-ray Imaging Polarimetry Explorer (XIPE)} mission, which 
has been proposed to ESA in response to a small-size (S-class) mission 
call due for launch in 2017.
\end{abstract}


\section{Introduction}

For the last two decades, an increasing number of type-1 active galactic nuclei (AGN) 
showing a broad Fe K$\alpha$ fluorescent line in the 4--7~keV band has been detected 
(see e.g. Reeves et al. 2006, Nandra et al. 2007, de la Calle et al. 2010, Patrick et al. 2011). 
The actual presence of the extended red wing of the line is confirmed in many objects; 
nonetheless, its physical origin is debated with two major interpretations emerging: 
a relativistic reflection scenario (Miniutti \& Fabian 2004) and an absorption scenario 
(Inoue \& Matsumoto 2003, Tatum et al. 2012).

The Seyfert galaxy MCG-6-30-15 is an archetypal case among AGN with broad iron lines. 
Its extended red wing is well-established from long observations with {\it XMM-Newton} 
(Wilms et al. 2001, Fabian et al. 2002) and {\it Suzaku} (Miniutti et al. 2007). 
Several authors interpret the line as reprocessed X-ray emission emerging from the accretion 
disc that reaches down to the innermost stable orbit (ISCO) of the supermassive black hole 
(Miniutti et al. 2003, Reynolds et al. 2009). In this view, the broadening is due to 
general relativistic and Doppler effects shifting the line centroid as a function of 
the disc radius. When integrating the emission across the whole disc while taking into account 
the effects of ray-tracing in a Kerr metric the line is ``relativistically blurred''. 
Assuming that the accretion disc and its irradiation are indeed truncated at the ISCO, 
the blurred line puts important constraints on the black hole spin 
(Fabian et al. 1989, Laor 1991, Dov{\v c}iak et al. 2004, Brenneman \& Reynolds 2006).

Following a different approach, Inoue \& Matsumoto (2003) and Miller, Turner \& Reeves (2008, 2009) 
argue that the X-ray data of MCG-6-30-15 can also be explained by assuming several absorbing media 
located on the line-of-sight and partially covering the primary X-ray source. In this interpretation, 
the extended red wing is ``carved out'' by absorption and the line shape is much less 
related to the SMBH spin.

More advanced spectral and timing analyses using forthcoming X-ray missions like {\it Astro-H} and {\it NuStar} 
in addition to {\it XMM-Newton} may shed more light on how broad iron lines are produced. 
In this letter, we still explore a different path and test how X-ray polarimetry can help to 
independently discriminate between the two models. 


\section[]{Comparison of the two scenarios}

The aim of this letter is to have a general view over the two competing scenarios. 
It is not the scope of this paper to produce an accurate spectral fit to the X-ray 
data of MCG-6-30-15. We rather assimilate prescriptions for reflection and absorption 
models that have been presented before and, based on these models, we compute the predicted 
X-ray polarisation as a function of the observer's viewing angle.

\begin{figure}
 \centering
 \includegraphics[trim =30mm 231mm 10mm 6mm, clip, width=13.5cm]{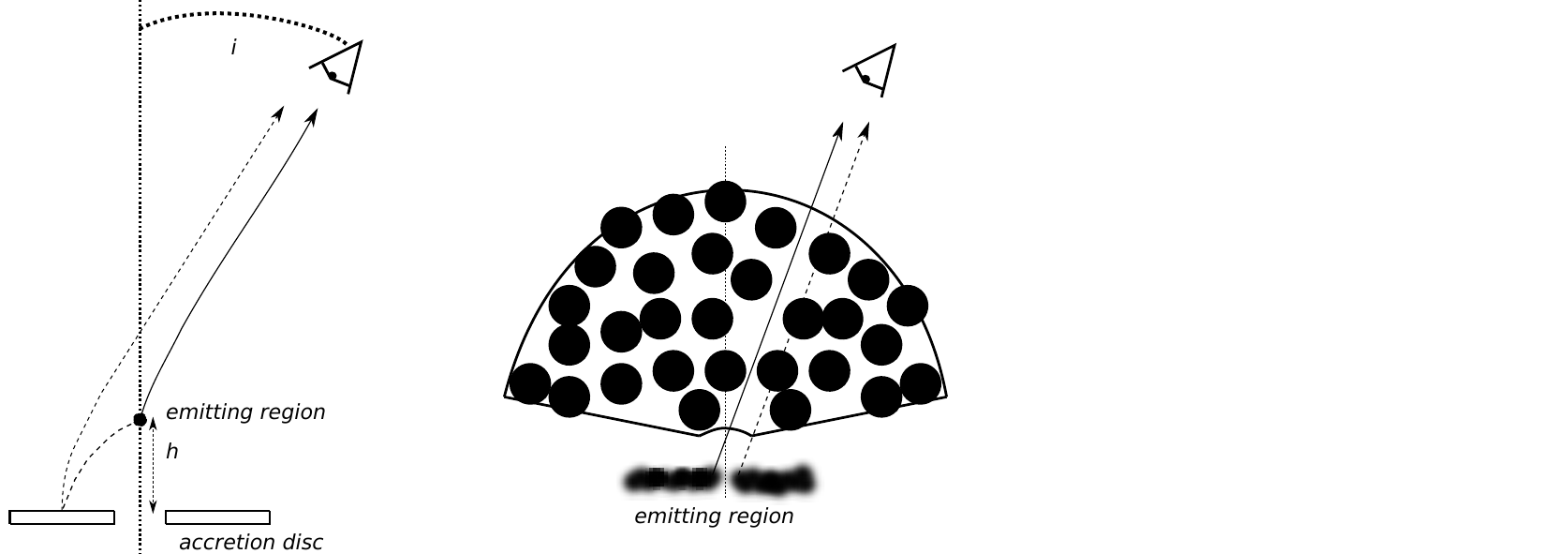}
 \caption{Schematic view of the scenarios considered. $Left$: reflection with a 
	  lamp-post geometry and light-bending. $Right$: partial covering with 
	  a clumpy wind.}
 \label{Fig0}%
\end{figure}

\subsection{The relativistic reflection model}
\label{sec:reflect}

We first consider relativistic reflection from a cold accretion disc illuminated by an 
elevated lamp-post on the disc axis. The method is described in detail in Dov{\v c}iak et al. (2011) 
so here we only give a brief summary. A grid of local reprocessing models, i.e. taken in the 
frame of the rotating accretion disc, was computed with the Monte-Carlo radiative transfer code 
{\it NOAR} (Dumont, Abrassart \& Collin 2000) providing the re-emitted intensity as a function 
of incident and re-emission angle. We defined an isotropic, point-like source emitting an 
unpolarised spectrum with a power law shape $F_{\rm *}~\propto~\nu^{-\alpha}$ and $\alpha = 1.0$.
The accretion disc is approximated by a constant density slab 
with cosmic abundances. Compton scattering, photo-absorption and iron line fluorescent emission 
are included in the computation of the locally re-emitted spectra. The local polarisation is 
computed using the transfer equations of Chandrasekhar (1960). Since the reprocessing 
medium is optically thick, the reprocessing predominately occurs very close to the irradiated 
surface of the slab and thus the approximation is sufficiently accurate. The local, polarised 
reflection spectra are then combined with the {\it KY}-code (Dov{\v c}iak, Karas \& Yaqoob 2004) 
conducting relativistic ray-tracing between the lamp-post, the disc, and the distant observer 
(see Fig.~\ref{Fig0}, left). The height of the lamp-post is fixed at 2.5~$R_{\rm G}$, where
 $R_{\rm G}= GM/c^2$, and an extreme Kerr black hole with the dimensionless 
spin $a = 1$ and a mass of $M = 1.5 \times 10^6 \rm M_\odot$ is assumed. 

Our choice of parameters is in good agreement with the assumptions of Miniutti \& Fabian (2004).
We point out though that in our approach the primary source is not off-axis, which should have an 
impact on the resulting polarisation. Models of a patchy corona 
(see e.g Galeev, Rosner \& Vaiana 1979, Haardt et al. 1994) presume that the off-axis sources should 
be anchored to the disc by magnetic field loops and thus co-rotate in Keplerian motion. In the 
relativistic reflection model, the X-rays are emitted very close to the black hole and we thus 
estimate the maximum orbital time-scale occurring in a corona with the radial size 
$R_{\rm C} = 50 R_{\rm G}$ to $317 \times \frac{M}{10^7 M_\odot} \left[ \left( \frac{R_{\rm C}}{R_{\rm G}} \right)^{1.5} + a \right] [s] \approx 17$~ks. 
This time span is by a large factor lower than the minimum exposure time for an observation of 
MCG-6-30-15 with a near-future X-ray polarimeter (see Sect.~\ref{sec:xipe}). The observed 
polarised flux due to a single, off-axis source is thus integrated over many Keplerian orbits. 
For this reason, the primary emission region should appear axis-symmetric in near-future X-ray 
polarimetry. So would the irradiation pattern due to a central lamp-post as assumed in our modelling.

The expected X-ray polarisation for a non axis-symmetric, clumpy irradiation pattern of the 
accretion disc was studied by Schnittman \& Krolik (2010), who obtain polarisation percentages 
across the 2--10~keV band that are significantly higher than the results obtained by Dov{\v c}iak et al. (2011) 
for the lamp-post geometry. Aside from the different coronal geometry, this is also related to different 
assumptions about the ionisation of the accretion disk. When assuming a radially 
structured surface ionisation (Ballantyne et al. 2003) we expect the local percentage of polarisation 
to increase compared to a neutral disc because the efficiency for electron scattering rises with ionisation. 
For the same black hole spin, the resulting integrated polarisation observed at infinity must therefore 
increase as well.

Finally, here we do not include any intrinsic polarisation of the primary radiation. 
Such polarisation may occur if the primary spectrum emitted by the lamp-post is indeed due to inverse 
Compton scattering of UV/X-ray photons coming from the disc. This effect may thus strengthen the net 
polarisation observed at infinity.

In summary, the net polarisation percentage predicted by our lamp-post model with unpolarised primary 
radiation and a cold accretion disk is likely to be conservative.

\begin{figure*}
 \centering
 \includegraphics[trim = 13mm 12mm 47mm 22mm, clip, width=17cm]{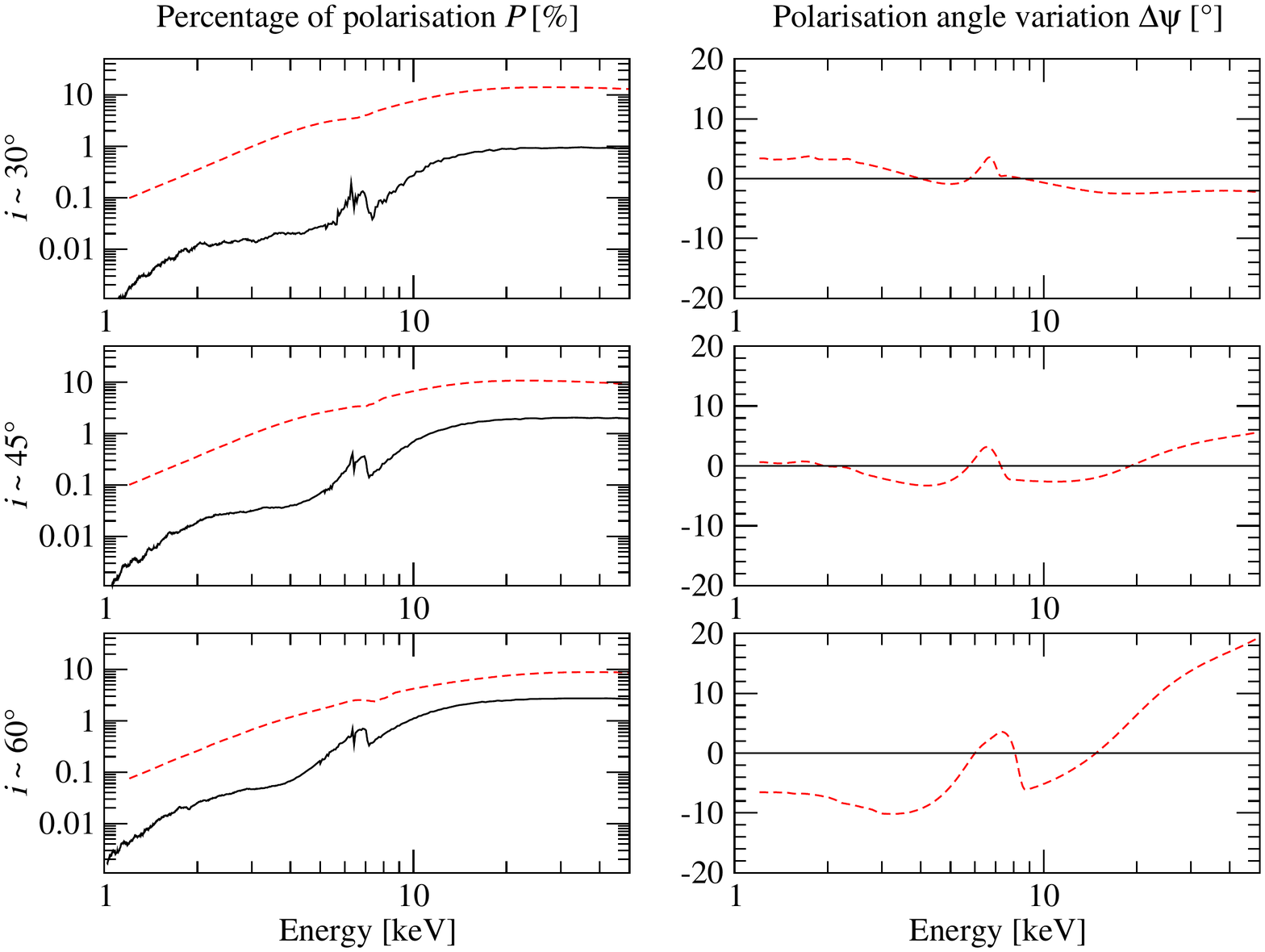}
 \caption{Percentage of polarisation $P$ and variation of the polarisation angle 
	  $\Delta\psi$ with respect to its mean as a function of the energy. 
	  Three particular viewing angles $i$ are considered : $30^\circ$, 
	  $45^\circ$ and $60^\circ$. $Legend$: a fragmented absorption region 
	  (plain line) and a relativistic reflection model with an extreme 
	  Kerr SMBH with $a=1$ (red dashed line).}
 \label{Fig2}%
\end{figure*}

\subsection{The complex absorption model}
\label{sec:abso}

An alternative approach to explain the broad red-wing of the Fe~K$\alpha$ line in MCG-6-30-15 
and its lack of variability with respect to the continuum was given by Miller et al. (2008). 
The authors first suggested a model of absorbed, non-relativistic reflection combined with 
variable partial covering of the primary source. In the following, Miller et al. (2009) even 
proposed a pure absorption scenario. This model supposedly is in-line with evidence for 
high-column density, partial covering absorption found in other AGN (Turner et al. 2009, 
Reeves et al. 2009, Risaliti et al. 2009). It contains five absorbing zones with the 
ionised zones 1--3 being required to reproduce the narrow absorption 
lines in the {\it Chandra} and {\it XMM-Newton} grating data. The spectral 
curvature in the 1--10~keV band is caused by the low-ionisation zones~4 and~5 covering the 
continuum source by 62\% and 17\%, respectively. We therefore focus on these two zones when 
modelling the expected polarisation. The absorbers 1--3 fully cover the source and they 
thus represent an additional, low optical depth of $0.0002 < \tau_{\rm c} < 0.06$ with 
respect to Compton scattering that we could add to zones~4 and 5. However, given the large 
optical depth of zone 4 ($\tau_{\rm c} \sim 1.5$) and its predominant covering factor, 
it turns out that the impact of zones~1--3 is very limited and that we can safely neglect them.

Using the latest version of the {\it STOKES} code (Goosmann \& Gaskell 2007, 
Marin et al., submitted), we model a geometrically thin, static, disc-like 
source emitting an isotropic, unpolarised primary spectrum between 1 and 100~keV 
using the same power law slope of $\alpha = 1.0$ as for the relativistic 
modelling presented in Sect. 2.1. This setup is close to the approach of Miller, Turner \& Reeves (2009). 
The emitting, central cylinder radially extends up to 0.05~pc and may represent 
a so-called hot inner flow. Note that in the absorption scenario we do not assume the accretion 
disc to reach down to the ISCO as otherwise, we would expect again the signs of relativistic 
reflection. The disk may be truncated at larger radii and thus it presents a low solid angle 
to the emission region and an eventual reprocessing component remains weak.

Between the source and the observer, a conical, neutral absorber with a height 
of 1.8~pc along the vertical axis, a half-opening angle of $75^\circ$, cosmic element abundances, 
and a Compton optical depth of either $\tau_{\rm c} \sim 1.5$ or $\tau_{\rm c} \sim 0.02$ 
is defined (Fig.\ref{Fig0}, right). This parameterisation is a good match to the modelling 
of zone~4 and zone~5 as given in Miller et al. (2009), except that our 
computations also include reprocessing. The actual modelling in {\it STOKES} is done for a 
uniform density cloud and the clumpiness is included by re-normalising the resulting Stokes 
fluxes in such a way that 62\% of the primary emission in the case of zone~4 
and 17\% in the case of zone~5 are incident onto the cloud while 21\% of the source flux 
reach the observer directly.

\subsection{Resulting polarisation signatures}

In Fig.~\ref{Fig2}, we plot the resulting polarisation as a function of photon energy 
at a viewing angle of $30^\circ$, $45^\circ$ and $60^\circ$. No circular polarisation can occur 
in this model setup, so $P$ designates only linear polarisation and $\Delta\psi$ the rotation 
of the polarisation position angle with respect to a convenient average of the polarisation 
position angles over the depicted energy band. The actual normalisation of the polarisation 
angle with respect to the disk axis is not of primary interest as we cannot determine it from 
the observations. 

It appears that in comparison with the absorption model, relativistic reflection produces a 
polarisation percentage $P$ in the 10-50 keV band that is at least by a factor 
of fifteen higher (Fig.~\ref{Fig2}, left). At lower energies, $P$ 
decreases gradually down to 0.1\% for the reflection scenario while 
for the absorption model $P$ drops much more drastically below 5 keV. 
In the relativistic case, the spectral shape of $P$ is determined by the net integration 
of the polarisation over the accretion disc. The fast motion of the accreting matter and 
strong gravity effects near the SMBH induce a rotation of the polarisation angle that 
depends on the position on the disc and on the inclination of the observer. In contrast 
to this, the energy dependence of $P$ for the absorption scenario is related to the 
polarisation phase function of electron scattering. A large fraction of the radiation 
has undergone mostly forward scattering and thus only produces weak polarisation. 

Additionally, the primary spectrum of the continuum source favours the emission of soft 
X-ray photons and thus it causes strong dilution of the transmitted flux by unpolarised 
radiation at low energies. In the reflection scenario, a significant part of the primary 
flux is bent down to the disk and the dilution is less efficient.

As the viewing angle increases from $30^\circ$ to $60^\circ$, $P$ varies differently in 
both scenarios. In the relativistic reflection case, $P$ decreases by a factor of $\sim$ 
1.3 as the polarisation contribution from the inner and outer parts of the disc produce 
differently oriented and partly cancelling polarisation. 
In the absorption case, $P$ increases with viewing angle; when taken between 5 and 8 keV 
it is by a factor of $\sim$ 6 higher at $60^\circ$ than at $30^\circ$. 
Nonetheless, the polarisation percentage for relativistic reflection always remains 
significantly higher than for the absorption scenario.

The variation of the polarisation angle (Fig.\ref{Fig2}, right) puts additional constraints 
on the origin of the broad iron line: in the absorption case, $\Delta\psi$ exhibits no variations. 
The relativistic model, however, induces energy-dependent variations in $\Delta\psi$ that 
increase with viewing angle and that are particularly strong across the iron line. 
This behaviour is related to the energy dependent albedo and scattering phase function of 
the disc material. Note that at a viewing angle of $60^\circ$ the variation of $\Delta\psi$ 
in the 2--10~keV band is larger than $10^\circ$. At an inclination of $30^\circ$, which is 
more probable for MCG-6-30-15, the variation is still around $5^\circ$.


\section{Observational prospects}
\label{sec:xipe}

The results plotted in Fig.~\ref{Fig2} show that, in principle, X-ray polarisation can 
distinguish between the relativistic reflection and the absorption for the broad iron line 
in MCG-6-30-15. Practically no polarisation signal should be measured in 
the 1--5 keV band for the absorption case and in the 10--50 keV band, the polarisation due 
to relativistic reflection is much higher than for the absorption scenario. The variation of 
the polarisation angle gives an additional handle on the scenario because the absorption model 
produces zero variation of $\Delta\psi$ while the reflection model induces smooth and 
characteristic variation of $\Delta\psi$ with photon energy.

\begin{figure}
 \centering
 \includegraphics[trim = 0mm 10mm 45mm 25mm, clip, width=8cm]{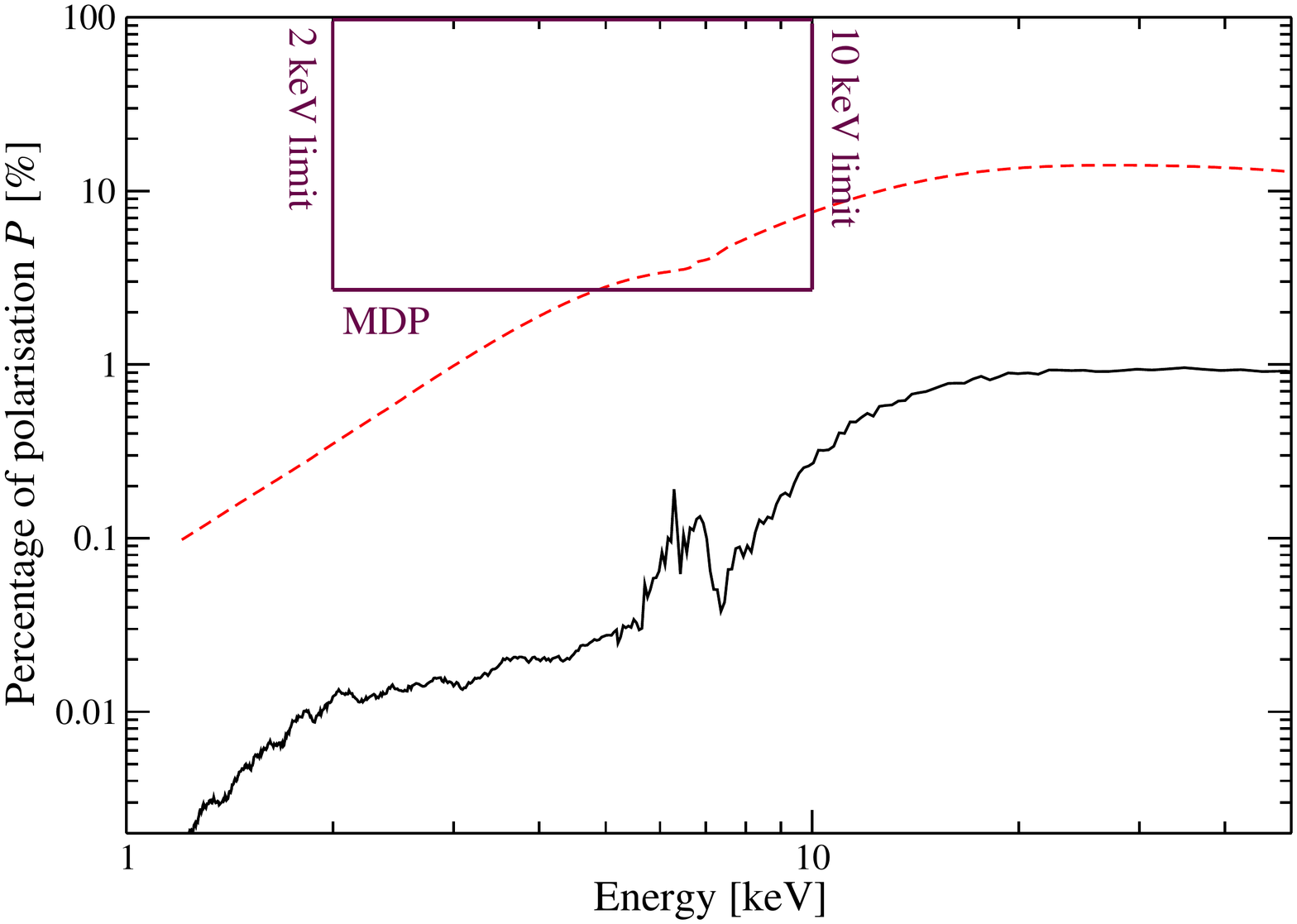}
 \caption{{\it XIPE} minimum detectable polarisation of the two scenarios for 
	  a 1~Ms observation of MCG-6-30-15 in the 2--10~keV band (maroon box). 
	  The observer's line-of-sight lies at $30^\circ$ with respect to the 
	  symmetry axis. {\it Legend:} clumpy absorption (plain line) and 
	  relativistic reflection induced by a Kerr SMBH with $a=1$ (dashed red line).}
 \label{Fig3}%
\end{figure}

A small X-ray polarimetry mission such as the {\it X-ray Imaging Polarimetry Explorer (XIPE)} 
being currently evaluated by the European Space Agency could already constrain the soft X-ray 
polarisation of MCG-6-30-15. The {\it XIPE} payload comprises the {\it Efficient X-ray Photoelectric 
Polarimeter (EXP)} dedicated to the observation of astrophysical sources in the 2--10 keV energy range. 
It has two Gas Pixel Detectors (Bellazzini et al. 2006, Bellazzini \& Muleri 2010) placed in the focal 
plane of two JET-X optics (Citterio et al. 1996). In Fig~\ref{Fig3}, we highlight the minimum detectable 
polarisation (MDP) of the {\it EXP} instrument at 99\% confidence level, which could be reached by a 
1~Ms observation of MCG-6-30-15 with a flux of 3~mCrab in the 2-10~keV band\footnote{When the 
background flux is negligible with respect to the source flux ($S$) the MDP of {\it XIPE} at a 
99\% confidence level is: 
MDP~$\approx 14\% \left( \frac{S}{1 \rm mCrab} \right)^{-1/2} \left( \frac{\rm exposure \, time}{100,000~\rm s} \right)^{-1/2}$.}. 
The polarisation expected for our reflection model is within the reach of {\it XIPE} and its detection would strongly support 
such a model. In case of a high significance detection, further indications could be derived by analysing the behaviour of 
the soft X-ray polarisation angle with energy.

The predictive power of our modelling would still greatly benefit from broad-band polarimetry 
as it was planned for the {\it New soft and Hard X-ray imaging and polarimetric Mission (NHXM)}. 
In the concept of this medium-sized mission, a 2--35 keV imaging polarimeter was included (Tagliaferri 2012). 
Such an instrument, which is technologically ready to fly today, would be very efficient to discriminate 
reflection from absorption also above 10~keV.

\section{Conclusions and future work}

The main result so far coming out of our modelling work is that with current observational technology 
the relativistic scenario should produce measurable soft X-ray polarisation while in the absorption 
case $P$ should be globally undetectable. If, in addition to that, $\Delta\psi$ can be determined to 
vary across the iron line, a second, independent indicator for the reflection scenario is found. 

The exact geometry of the absorber situated along the observer's line-of-sight is unconstrained. To 
further support the results presented in this paper, we currently explore a range of 
different absorption scenarios and physical properties of the outflow, and we are going to present 
their polarisation characteristics in future work. Our modelling of the fragmented medium will be 
refined by using randomly-situated, spherical absorbers of constant density along the observer's 
line-of-sight. While being more realistic, such a configuration is expected to produce even lower 
polarisation percentage than the absorption model presented here. In a clumpy medium, the radiation 
has to undergo multiple scattering events that have a depolarising effect. 

We have also started to investigate the polarisation expected from a wind geometry such as the one 
suggested by Elvis (2000). In this scheme, the absorbing wind arises vertically from a narrow range of 
radii on the accretion disc and due to radiation pressure it is bent outward in a conical shape. 
Similar wind geometries were investigated by Sim et al. (2008,2010) and Schurch et al. (2009)
from hydrodynamic simulations. For a distant observer looking at the far-end of the wind, the system is 
seen in absorption. Preliminary tests show that such wind models produce polarisation that is slightly 
different from the one obtained for the conical outflows described in Sect.~\ref{sec:abso} but the results 
remain within the margins of our conclusions. Note that a partially ionised absorber 
may produce a stronger reprocessing component than a neutral wind. However, since forward-scattering 
predominates, the net polarisation is again expected to be low but it may depend on the viewing angle, 
the geometry of the medium or its ionisation structure. We are going to investigate such scenarios in 
more detail imposing that they correctly reproduce the observed broad spectral shape of the iron line.

It is also necessary to look into more realisations of the relativistic reflection scenario. 
In Sect.~\ref{sec:reflect}, we argue why changes in the irradiation geometry, the ionisation of the 
accretion disc or the polarisation of the primary radiation with respect to the current model should 
lead to stronger polarisation. We still need to verify this assumption by adopting the 
radial ionisation profile used in Svoboda et al. (2012), local reprocessing computations for ionised 
media and intrinsically polarised X-ray emission.

For now, we summarise our conclusion as follows: if a small, soft X-ray polarimetry mission like {\it XIPE} 
observes MCG-6-30-15 and detects polarisation in the 2--10~keV band the reflection scenario is confirmed. 
If there is no detection of polarisation the absorption scenario is more likely to be correct but a 
complex reflection model cannot be excluded.

\section*{Acknowledgments}

We thank the anonymous referee for helpful comments. This research was supported by 
the grants ANR-11-JS56-013-01, COST-CZ LD12010, the GdR PCHE, and the exchange program CNRS/Academy 
of Sciences of the Czech Republic.


\bsp

\label{lastpage}

\end{document}